\documentclass[11pt]{article}
\usepackage{epsf}
\usepackage{cite}
\newcommand{\floatwidth}{4.5in}

\begin{document}
\bibliographystyle{prsty} 

\begin{flushright}
SU-4240-629\\
SCCS-754\\
NBI-HE-96-12\\
March 1996
\end{flushright}

\begin{center}
\vspace{24pt}

{\Large \bf The Flat Phase of Crystalline Membranes}
\vspace{24pt}

{\large \sl Mark J. Bowick, Simon M. Catterall} \\
{\large \sl Marco Falcioni and Gudmar Thorleifsson}
\vspace{6pt}

Department of Physics \\
Syracuse University \\
Syracuse, NY 13244-1130, USA\\
{\tt http://www.phy.syr.edu/}
\vspace{10pt}

{\large \sl Konstantinos N. Anagnostopoulos} \\
\vspace{6pt}

The Niels Bohr Institute \\
Blegdamsvej 17\\
DK-2100 Copenhagen \O , Denmark \\
{\tt http://www.nbi.dk/}
\vspace{10pt}

\begin{abstract}
\noindent We present the results of a high-statistics Monte Carlo
simulation of a phantom crystalline (fixed-connectivity) membrane with
free boundary.  We verify the existence of a flat phase by examining
lattices of size up to $128^2$.  The Hamiltonian of the model is the
sum of a simple spring pair potential, with no hard-core repulsion,
and bending energy.  The only free parameter is the the bending
rigidity $\kappa$.  In-plane elastic constants are not explicitly
introduced.  We obtain the remarkable result that this simple model
dynamically generates the elastic constants required to stabilise the
flat phase.  We present measurements of the size (Flory) exponent
$\nu$ and the roughness exponent $\zeta$.  We also determine the
critical exponents $\eta$ and $\eta_u$ describing the scale dependence
of the bending rigidity ($\kappa(q) \sim q^{-\eta}$) and the induced
elastic constants ($\lambda(q) \sim \mu(q) \sim q^{\eta_u}$). At
bending rigidity $\kappa = 1.1$, we find $\nu = 0.95(5)$ (Hausdorff
dimension $d_H = 2/\nu = 2.1(1)$), $\zeta = 0.64(2)$ and $\eta_u =
0.50(1)$.  These results are consistent with the scaling relation
$\zeta = (2+\eta_u)/4$.  The additional scaling relation $\eta =
2(1-\zeta)$ implies $\eta = 0.72(4)$. A direct measurement of $\eta$
from the power-law decay of the normal-normal correlation function
yields $\eta \approx 0.6$ on the $128^2$ lattice.
\end{abstract}
\end{center}
\newpage
\section{Introduction}
\label{sec:intro}
The physics of flexible membranes, two-dimensional surfaces
fluctuating in three dimensions, is extremely rich, both on the
theoretical side, where there is a nice interplay between geometry,
statistical mechanics and field theory, and on the experimental side,
where model systems abound.

The simplest examples of 2-dimensional surfaces are monolayers, or
films---these are strictly planar systems.  The statistical behaviour
of monolayers falls into three distinct universality classes,
depending on the microscopic interactions of the system. There are
{\it crystalline} monolayers, with quasi-long-range translational
order and long-range orientational order, {\it hexatic} monolayers,
with short-range translational order but quasi-long-range
orientational order, and {\it fluid} monolayers, with short-range
translational and orientational order \cite{DRNLH}.

Physical membranes, 2-dimensional surfaces fluctuating in a
3-dimensional embedding space, are expected to have an equally rich
phase structure.  The simplest class of membranes is the {\it
crystalline} class, in which topological defects, dislocations, and
disclinations, are forbidden.  At the discrete level crystalline
membranes may be modelled by triangulated surfaces with fixed local
connectivity.  In this paper we will be concerned with the properties
of a particularly simple model of a phantom (non self-avoiding)
crystalline membrane, with emphasis on a critical analysis of the
existence and stability of an ordered {\em flat} phase of the model
for sufficiently large {\em bending rigidity}.

Crystalline membranes, also known as tethered or polymerised
membranes, are the natural generalisation of linear polymer chains to
two dimensions.  Polymer chains in a good solvent crumple into a
fractal object with Hausdorff dimension 5/3 (Flory exponent
$\nu=3/5$).  Crystalline membranes with in-plane solid-like
elasticity, on the other hand, are predicted to exhibit quite
different physical properties from their linear counterparts.  In
particular, they are expected to have a remarkable low-temperature
ordered phase.  This ordered, or {\em flat}, phase is characterised by
long-range order in the orientation of surface normals.  At high
temperature, or equivalently low bending rigidity, phantom crystalline
membranes will entropically disorder and crumple.  Separating these
two phases should be a crumpling transition, whose precise nature for
physical membranes is still not fully understood.

Inorganic examples of crystalline membranes are thin sheets ($\le 100$
\AA) of graphite oxide (GO) in an aqueous suspension
\cite{graphite1,graphite2} and the rag-like structures found in
MoS$_2$ \cite{CPPDN}.

There are also beautiful biological examples of crystalline membranes
such as the spectrin skeleton of red blood cell membranes. This is a
two-dimensional triangulated network of roughly 70,000
plaquettes\footnote{The $128^2$ lattice we simulate has 32,766
plaquettes.}.  Actin oligomers form nodes and spectrin tetramers form
links \cite{Skel}. Crystalline membranes can also be synthetised in
the laboratory by polymerising amphiphillic monolayers or
bilayers. For recent reviews see \cite{Peliti,fdavid,jer2}.

The existence of an ordered phase in a two dimensional system is
remarkable, given the Mermin-Wagner theorem.  In fact, if one thinks
of surface normals as spins, then the membrane bending energy is akin
to a Heisenberg interaction, and the $2d$-Heisenberg model is well
known to have no ordered phase.  What stabilises the flat phase in a
crystalline membrane?  There are several appealing arguments for the
existence of a stable ordered phase in crystalline membranes
\cite{NP1}.  Out-of-plane fluctuations of the membrane are coupled to
in-plane ``phonon'' degrees of freedom because of the non-vanishing
elastic moduli (shear and compressional) of a polymerised membrane.
Bending of the membrane is inevitably accompanied by an internal
stretching of the membrane.  Integrating out the phonon degrees of
freedom one finds long-range interactions between Gaussian curvature
fluctuations which stabilise a flat phase for sufficiently large bare
bending rigidity\cite{NP1}.  Alternatively both the elastic constants
and the bending rigidity pick up anomalous dimensions in the field
theory sense. The bending rigidity receives a stiffening contribution
at large distances via the phonon coupling.  This competes with the
usual softening of bending rigidity seen in fluid membranes, with the
net result being an ultraviolet-stable fixed point---the crumpling
transition.  From the magnetic point of view membrane models are
constrained spin systems, since the spins must be normal to the
underlying surface, and the constraints are essential to the stability
of the ordered phase.

This viewpoint is supported by self-consistent calculations for the
renormalisation of the bending rigidity, by large-$d$ calculations,
where $d$ is the dimension of the embedding space, and by
$\epsilon$-expansion calculations, where $\epsilon = 4-D$, with the
internal dimension $D$ of the membrane being 2 for physical
polymerised membranes.

The construction of a discrete formulation of a crystalline membrane
is essential for numerical simulations, and revealing for comparison
with spin systems.  In the simplest discretised version the membrane
is modelled by a regular triangular lattice with {\it fixed}
connectivity, embedded in $d$-dimensional space.  Typically the
link-lengths of the lattice are allowed some limited fluctuations.
This may be modelled by tethers between hard spheres or by introducing
some confining pair potential with short-range repulsion between nodes
(monomers).  The bending energy is represented by a ferromagnetic-like
interaction between the normals to nearest-neighbour ``plaquettes''
\cite{jer2}.
 
In the random surface literature, on the other hand, it is more common
bto consider a simple Gaussian spring model of the pair potential with
vanishing equilibrium spring length.  In this case the minimum
link-lengths are unconstrained.  A priori this model seems to be
rather different from those above; one may worry that it is
pathological in some sense.  In particular one may ask if this model
can ever generate the effective elastic constants which are required
to stabilize a flat phase at large bending rigidity.  To illustrate
this concern consider the infinite bending rigidity limit of the
model. In this limit the system becomes planar --- {out-of-plane}
fluctuations are completety suppressed.  Since the pair potential
allows it, what prevents all nodes from collapsing on each other
within the plane of the membrane itself?  The answer can be seen by
considering what happens if a node moves across a bond, or if two
nodes are exchanged.  In this case the normals to some triangular
plaquettes will inevitably be inverted. This generates a prohibitive
bending energy cost and is effectively forbidden. The model thus
dynamically generates a hard-core repulsion\footnote{We thank one of
the referees for informative observations on this point}, and may be
thought of as having an effective equilibrium spring length.  In
Appendix A we show that small fluctuations about a finite microscopic
equilibrium spring length $a$ yield elastic constants proportional to
${(a/\epsilon)}^2$, where $\epsilon$ is the intrinsic lattice spacing.
For $\epsilon \simeq a$ these constants are finite.  For the model we
consider, with vanishing microscopic $a$ and finite $\epsilon$, the
heuristic arguments above suggest one should replace $a$ by an
effective equilibrium spring length.

Monte Carlo simulations have in fact established strong evidence for a
continuous crumpling transition in the model with $a=0$. Most of the
simulations, however, have focused on the crumpling transition itself.
Not much effort has been made to establish rigorously the existence
and the properties of the flat phase.  In the rest of this paper we
present evidence that there is indeed a stable flat phase in this
remarkably simple model of a crystalline membrane. Furthermore we 
show that the requisite elastic constants are dynamically generated
with the correct scaling behaviour.

The paper is organised as follows: in Section \ref{sec:model} we
discuss the discrete model, the numerical simulations and the choice
of the boundary conditions.  In Section \ref{sec:crumple} we review
the evidence for the crumpling transition.  In Section \ref{sec:flat}
we discuss the Monge representation of a surface and the theoretical
predictions for asymptotically flat elastic surfaces.  We then discuss
our numerical results for the roughness exponent, the phonon
fluctuations and the normal-normal correlation function.  In Appendix
A we give a calculation of the elastic constants of a discrete
soft-core Gaussian model. In Appendix B we discuss of the methods used
to measure the geodesic distance for the correlation function and in
Appendix C we describe the parallel algorithm used to simulate the
lattices with largest size.
\section{Model}
\label{sec:model}
\begin{figure}
\centerline{\epsfxsize=4.7in \epsfbox{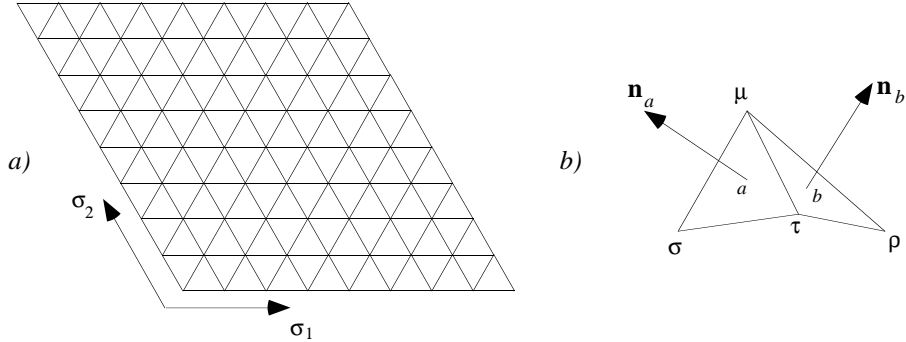}}
\caption{{\it a)} The intrinsic connectivity structure of the
mesh. {\it b)} The labelling scheme: $\sigma$, $\rho$, $\tau$, $\mu$ 
are the
intrinsic coordinates while $a$, $b$ label the triangles.}
\label{fig:mesh}
\end{figure}
To describe a discrete polymerised surface we arrange $N$ particles
(monomers) in a regular triangulation of a 2--$D$ manifold (see Fig.\
\ref{fig:mesh}{\it a}).  The 2--$D$ surface is then embedded in a
$d$--dimensional space where it is allowed to fluctuate in all
directions.  Each monomer is labelled by a set of intrinsic
coordinates ${\bf \sigma} = (\sigma_1,\sigma_2)$, with respect to a
set of orthogonal axes in the 2--$D$ manifold.  The position in the
$d$--dimensional embedding space is given by the vector ${\bf
x}_{\sigma}$.  We will treat the case $d = 3$.

In general, the Hamiltonian which describes these models has two
terms: a pair potential and a bending energy term,
\begin{equation}
\label{eqn:generalaction}
{\cal H} = {\cal H}_T + {\cal H}_B.
\end{equation}
A commonly studied pair potential in the literature
\cite{KN1} is the tethering potential with hard core repulsion, 
given by
\begin{equation}
\label{eqn:hard-core}
{\cal H}_{T} = \sum_{\langle {\bf \sigma} {\rm {\sigma^\prime}} 
\rangle}
V(\vert {\bf x}_{\sigma} - {\bf x}_{\sigma^\prime} \vert),
\end{equation}
with
\begin{equation}
V(R) = \left\{
\begin{array}{cl} 
\infty & R < a\\ 0 & a \le R \le b\\ \infty & R > b
\end{array}\right. , \; R = \vert  {\bf x}_{\sigma} -  {\bf
 x}_{\sigma^\prime} \vert
\end{equation}
where $R$ is the link length, $a$ is the hard-core radius and $b$ is
the tethering length.

We consider, instead, a model where the tethering potential is
replaced by a simple Gaussian spring potential. The bending energy
is the usual ferromagnetic interaction between the normals to the
faces of the membrane,  namely
\begin{equation}
\label{eqn:ouraction}
{\cal H} = \sum_{\langle \sigma \sigma^\prime \rangle} \left| {\bf
x}_{\sigma} - {\bf x}_{\sigma^\prime}\right|^2 + \kappa \sum_{\langle 
a b
\rangle} \left( 1 - {\bf n}_{a} \cdot {\bf n}_{b}\right).
\end{equation}
Here $\kappa$ is the bending rigidity, $a$ and $b$ label the faces of
the surface and ${\bf n}_a$ is the unit normal to the face $a$.  Both
sums extend over nearest neighbours (see Fig.\ \ref{fig:mesh}{\it
b}). This action has no explicit short scale cut-off length or
hard-core repulsion.  Since there is no self-avoidance it represents a
{\em phantom surface}.  This action was investigated originally by
Ambj{\o}rn {\em et al}.~\cite{ADJ} and is inspired by the Polyakov
action for Euclidean strings with extrinsic curvature
\cite{Klein,Poly}.

The main justification for studying phantom surfaces as models of
realistic membranes is that in the flat phase self-avoidance is
irrelevant \cite{KN1,KK}. Although self-avoidance is expected to
change the scaling behaviour at the critical point $\kappa_c$ and the
nature of crumpled phase, it {\em does not} affect the scaling
behaviour in the flat phase.  As membranes with strong self-avoidance
(and the resultant non-local interactions) are harder to simulate
numerically, it is sensible to leave out self-avoidance when possible.

The partition function of this model is given by the trace of the
Boltzmann weight over all possible configurations of the embedding
variables ${\bf x}$:
\begin{equation}
\label{eqn:partfun}
{\cal Z} = \int [d{\bf x}] ~ \delta ({\bf x}_{cm}) \; \exp( -
 {\cal H}[{\bf x}]).
\end{equation}
Here ${\bf x}_{cm}$ is the centre of mass, which is held fixed to
eliminate the translational zero mode.  Expectation values are given
by
\begin{equation}
\label{eqn:expect}
\langle {\cal O} \rangle = \frac{1}{\cal Z} \int [d{\bf x}] ~ \delta
 ({\bf x}_{cm}) {\cal O}[{\bf x}] \exp( - {\cal H}[{\bf x}]).
\end{equation}
We will consider the case of surfaces with the topology of the disk
and free boundary conditions.  Most experimental realisations of
membranes have either this topology or spherical topology. Our choice
also has certain technical advantages.  To describe the flat phase we
need to measure the finite-size-scaling of the thickness of the
surface and the asymptotic behaviour of the normal-normal correlation
function.  With spherical or toroidal topology one would have to
subtract the effects of the global shape of the surface.
\subsection{Numerical Simulations}
To evaluate the integral of Eq.\ (\ref{eqn:expect}) we use the Monte
Carlo algorithm with a Metropolis update.  In our case the Metropolis
update corresponds to changing the position of a node by a trial
vector $\vec{\epsilon}$ chosen randomly (and uniformly) in a box of
size $(2\delta)^3$ centred on the old node position.  The update is
accepted if the change in the Hamiltonian is such that
\begin{equation}
\exp ({\cal H}_{old} - {\cal H}_{new}) > r,
\end{equation}
where $r$ is a uniformly distributed random variable with values in
the interval $[0,1)$.  The value of $\delta$ is adjusted to keep the
acceptance ratio around 50\%.  For a surface of linear size $L$, one
Monte Carlo sweep corresponds to Metropolis update of all $L^2$ nodes.
We used a lagged Fibonacci pseudo-random number generator.
\begin{table}
\begin{center}
\begin{tabular}{|l|c|c|c|}\hline
$L$ & $\kappa = 1.1$ & $\kappa = 2.0$ & $\tau_R \sim L^z$ \\ \hline 
32 & $31 \times 10^6$ & $26 \times 10^6$ & $3 \times 10^4$ \\ 
46 & $51 \times 10^6$ & $42 \times 10^6$ & $7 \times 10^4$ \\ 
64 & $47 \times 10^6$ & $44 \times 10^6$ & $1.2 \times 10^5$\\
128 & $74 \times 10^6$ & --- & $5 \times 10^5$\\ \hline
\end{tabular}
\end{center}
\caption{The number of thermalised sweeps collected per data point in
the flat phase. The last column indicates the autocorrelation time
for the slowest mode in the system, the radius of gyration.
The autocorrelation time is comparable for both values of $\kappa$.
The critical slowing down exponent $z \approx 2$, as
expected for a local algorithm.}
\label{tbl:runs}
\end{table}
For the simulations we used both scalar and parallel machines: a
MASPAR MP1 massively parallel processor, a 12-node IBM SP2 and an
8-node DEC Alpha farm.  The SP2 and the Alphas were used as single
independent CPUs, but we used a parallel Monte Carlo algorithm for the
simulation of the largest lattice ($L=128$) on the MP1. We show in
Table \ref{tbl:runs} the amount of thermalised data collected for the
various lattice sizes and couplings in the flat phase. For the largest
lattice we thermalised our surfaces by discarding about $10^7$
sweeps. In addition we have performed simulations close to the
crumpling transition. This work is still in progress but preliminary
results are discussed in Section \ref{sec:crumple}.

Previous studies of the critical behaviour of crystalline surfaces
used more elaborate simulation algorithms which combine a Langevin
update with Fast Fourier Acceleration
\cite{BEW,HW,WS,BET,ET1,ET2,Wheater}.  This algorithm is known to
be more effective in reducing critical slowing down. We chose not to
use this approach for several reasons.  First of all it is very hard
to implement the Fast Fourier Transform (FFT) on a 2-dimensional
surface with free boundaries.  Secondly the parallel computer we are
using is less efficient on floating point intensive algorithms, such
as Fourier Transforms.  Finally, the Langevin algorithm suffers from
systematic errors induced by the finite time step $\Delta t$.  This
necessitates an extrapolation to $\Delta t = 0$, which can itself
become time consuming.

Our statistical errors were estimated using two methods: the first is
a direct measurement of the autocorrelation in the data and a
corresponding correction of the standard deviation. The second is the
standard jackknife algorithm.  Both methods give consistent results.

\subsection{Boundary Effects}
\label{sec:boundary}
\begin{figure}
\centerline{\epsfxsize=3.9in \epsfbox{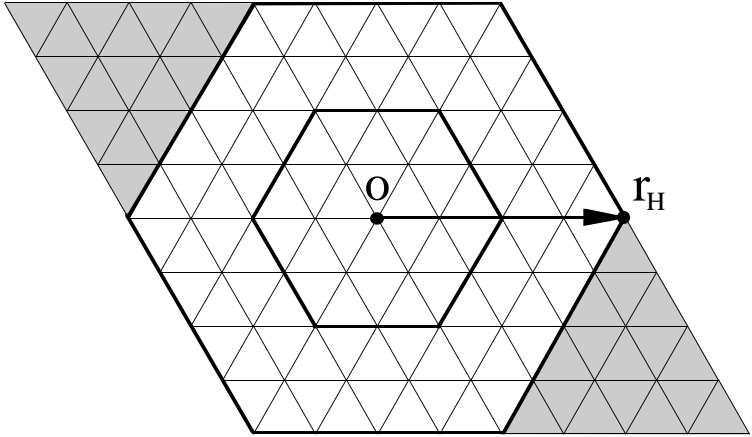}} 
\caption{Two
of the hexagons used to define the averages of the integrated
observables.  The arrow indicates the radius of the largest hexagon.}
\label{fig:hexagons}
\end{figure}

The optimal shape for a triangulated surface with free boundaries is a
hexagon. In our simulations the use of the parallel computer MP1 makes
the hexagonal mesh inconvenient, since the layout of the CPUs is a
square grid.  When we map the regular triangulation to the square
grid, the resulting surface has a rhomboidal shape, as can be seen
from Fig.\ \ref{fig:hexagons}.  In particular the regions in the
shaded areas of Fig.\ \ref{fig:hexagons} will be strongly influenced
by the boundaries.
For a generic observable ${\cal O}_\sigma$ we want to be able to
quantify the effect of the boundary and of the anisotropy.  In order
to achieve this we integrate the observable over hexagonal subsets of
the mesh, centred with respect to the surface.  Fig.\
\ref{fig:hexagons} shows two of these subsets with darker lines.  For
a surface of linear size $L$ we construct $L/2$ such integrated
observables by
\begin{equation}
\bar{\cal O}_r = {1 \over N_r} \sum_{\sigma \in H_r} {\cal O}_\sigma
\end{equation}
where $H_r$ is a hexagon of radius $r$ and $N_r$ the appropriate
normalisation.  The shaded areas of Fig.\ \ref{fig:hexagons} are
discarded from the integration. By looking at larger and larger
hexagons we can see when the boundaries start to affect the data.  For
very small hexagons the discretisation effects are large.  In practice
we find that the results are strongly influenced by the boundary for
hexagons of radius $r > L/4$.

For non-integrated observables, such as the normal-normal correlation
function, translational invariance {\em in the surface} is broken by
the presence of the free boundary.  Thus we always consider the
correlation function from the centre of the surface to all the other
nodes.  The effect of the boundary can then be seen clearly on the
correlations at a distance of order $L/4$.  This boundary data is
discarded from the fits.

\section{Crumpling Transition}
\label{sec:crumple}
Before examining the flat phase of the model of Eq.\
(\ref{eqn:ouraction}) we would like to review the existing evidence
for a ``crumpling'' transition.  In recent years the crumpling
transition has been the focus of extensive numerical and analytical
investigation.  Within the condensed-matter community it is customary
to work with models with effective potentials like Eq.\
(\ref{eqn:hard-core}) and free boundaries \cite{KKN1,KN1,KN2}.
Numerical simulations of these models provide direct evidence for a
phase transition, such as a diverging specific heat, although the
accuracy is not yet sufficient for a reliable determination of the
specific heat exponent $\alpha$.  When strong self-avoidance is
included in the models described by Eq.\ (\ref{eqn:hard-core}), there
is numerical evidence that the crumpling transition disappears
altogether, with the system being flat for all bending rigidities
\cite{PB,BLLP,ARP,HB,GM}.  Some studies of flexible impenetrable
plaquettes, however, do find a crumpled phase \cite{B,BR,KG}.  The
Gaussian spring models have been studied numerically in
\cite{BEW,HW,WS,BET,ET1,ET2,ADJ,Wheater,KR} using periodic boundary
conditions, with emphasis on the precise nature of the phase
transition. A growing peak in the specific heat is observed and the
best estimate of the related critical exponent is $\alpha = 0.58(10)$
\cite{Wheater}.

There is thus strong evidence that models of phantom polymerised
membranes have a continuous phase transition.  We are also currently
investigating this transition, and in the rest of this section we
discuss our preliminary results.
\subsection{Specific Heat}
\label{sec:specheat}
\begin{figure}
\centerline{\epsfxsize= \floatwidth \epsfbox{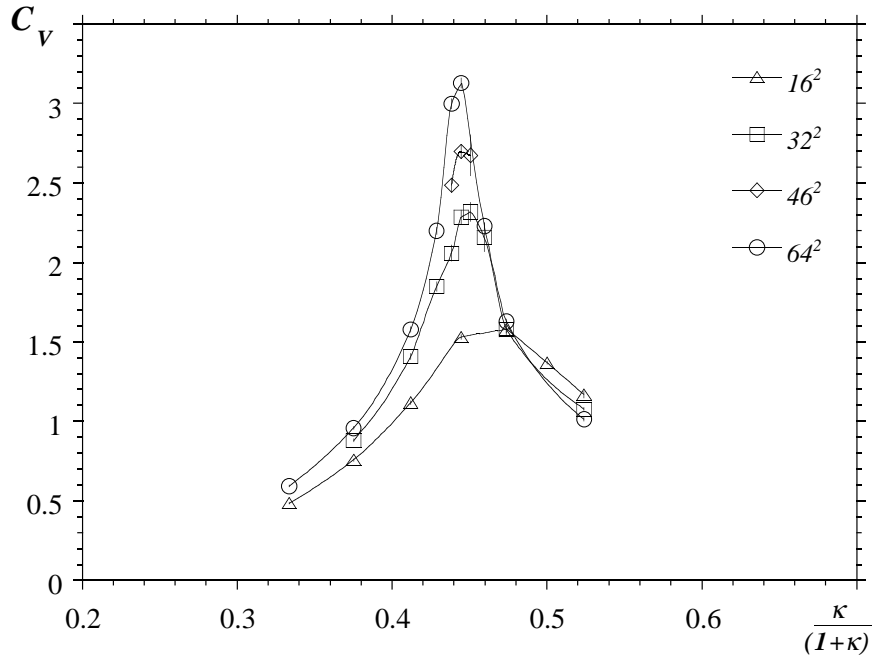}}
\caption{The specific heat versus bending rigidity.}
\label{fig:specplot} 
\end{figure}
Let us now turn to the energy fluctuations in our model. We write the
total bending energy as
\begin{equation}
\label{eqn:extcurv}
S_e = \sum_{\langle a b \rangle} {\bf n}_a \cdot {\bf n}_b.
\end{equation}
Denoting the number of links (or edges) in the surface by $N_e$,
it is simple to show \cite{HW} that the specific heat is given by
\begin{equation}
\label{eqn:specheat}
C_V = \frac{3(N-1)}{2} + \frac{\kappa^{2}}{N_e} ( \langle S_e^2 
\rangle
- \langle S_e \rangle^2 ).
\end{equation}
Here the brackets indicate a statistical average over
surfaces\footnote{As our statistics close to the phase transition are
not as good as in the flat phase, we have not attempted to use the
more sophisticated method described in Section \ref{sec:boundary} to
eliminate boundary effects from this data.}.  We henceforth drop the
constant piece from our analysis.

We plot in Fig.\ \ref{fig:specplot} the measured specific heat versus
bending rigidity, for surfaces consisting of up to $64^2$ nodes.  As
expected we see a sharp peak at $\kappa \approx 0.79$ growing with
system size.  The critical behaviour of $C_V$ close to the phase
transition is governed by an exponent $\alpha$, $C_V \sim \vert \kappa
- \kappa_c \vert^{-\alpha}$, and for $\alpha < 1$ the phase transition
is continuous (as the first derivative of the free energy does not
diverge).  Hence it is important to determine the value of $\alpha$.
The most convenient way of doing so is using finite size scaling,
which predicts that the value of the peak should scale with volume as
\begin{equation}
\label{eqn:alphadef}
C_V \; \approx \; c_0 + c_1 L^{\omega},
\end{equation}
where, assuming hyperscaling, the specific heat exponent $\alpha = {2
\omega\over 2 + \omega}$ and $c_0$ and $c_1$ are non-universal
constants.  Our best estimate of $\omega$, from the data shown in
Fig.\ \ref{fig:specplot}, is $\omega = 0.5(1)$, consistent with
previous results \cite{Wheater,KR}.  The corresponding value of
$\alpha$ is $0.4(1)$.
\begin{figure}
\centerline{ \epsfxsize= \floatwidth \epsfbox{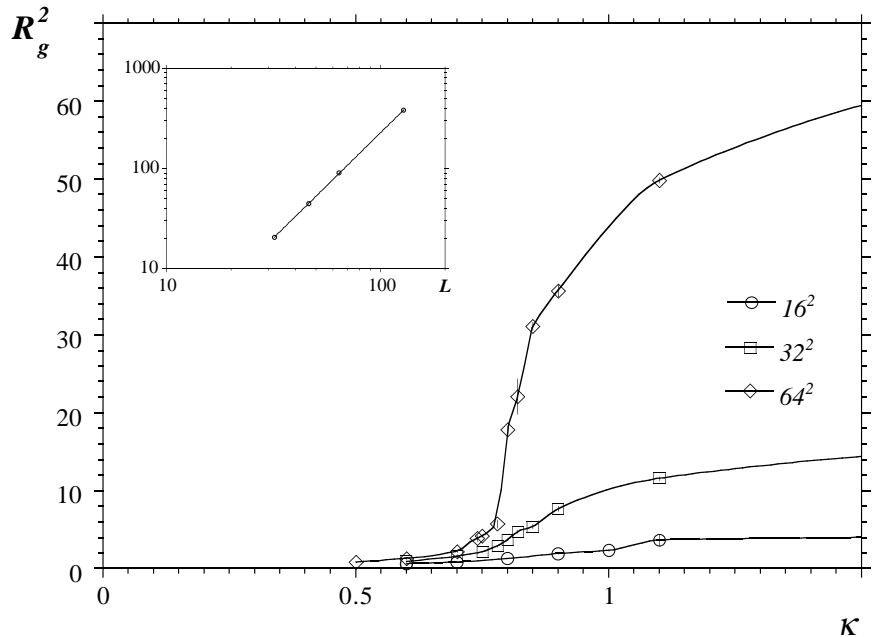}}
\protect\caption{The radius of gyration as a function of $\kappa$ for
various system sizes.  In the inset we show a power-law fit for $R_g$
vs.\ $L$ at $\kappa = 1.1$ for $L$ up to 128. We get $\nu = 0.95(5)$.}
\protect\label{fig:rg}
\end{figure}
\subsection{Radius of Gyration}
\label{sec:rg}
While the specific heat peak is one signal for the existence of a
phase transition, it reveals little about the nature of the phases
each side of the transition.  An observable more sensitive to the
geometry of the surface is the square of the {\it radius of gyration},
\begin{equation}
\label{eqn:gyra}
R^2_g = \left\langle \frac{1}{3N} \sum_{\sigma} {\bf x}_{\sigma} \cdot
{\bf x}_{\sigma} \right\rangle,
\end{equation}
which measures the average spatial extent of the surface.  $R_g$
defines a linear length scale for the surface and can be used to define
a fractal or Hausdorff dimension $d_H$ in the embedding space via
\begin{equation}
\label{eqn:hausdef}
R_g \sim N^{1/d_H}.
\end{equation}
\begin{figure}
\centerline{ \epsfxsize=4.7in \epsfbox{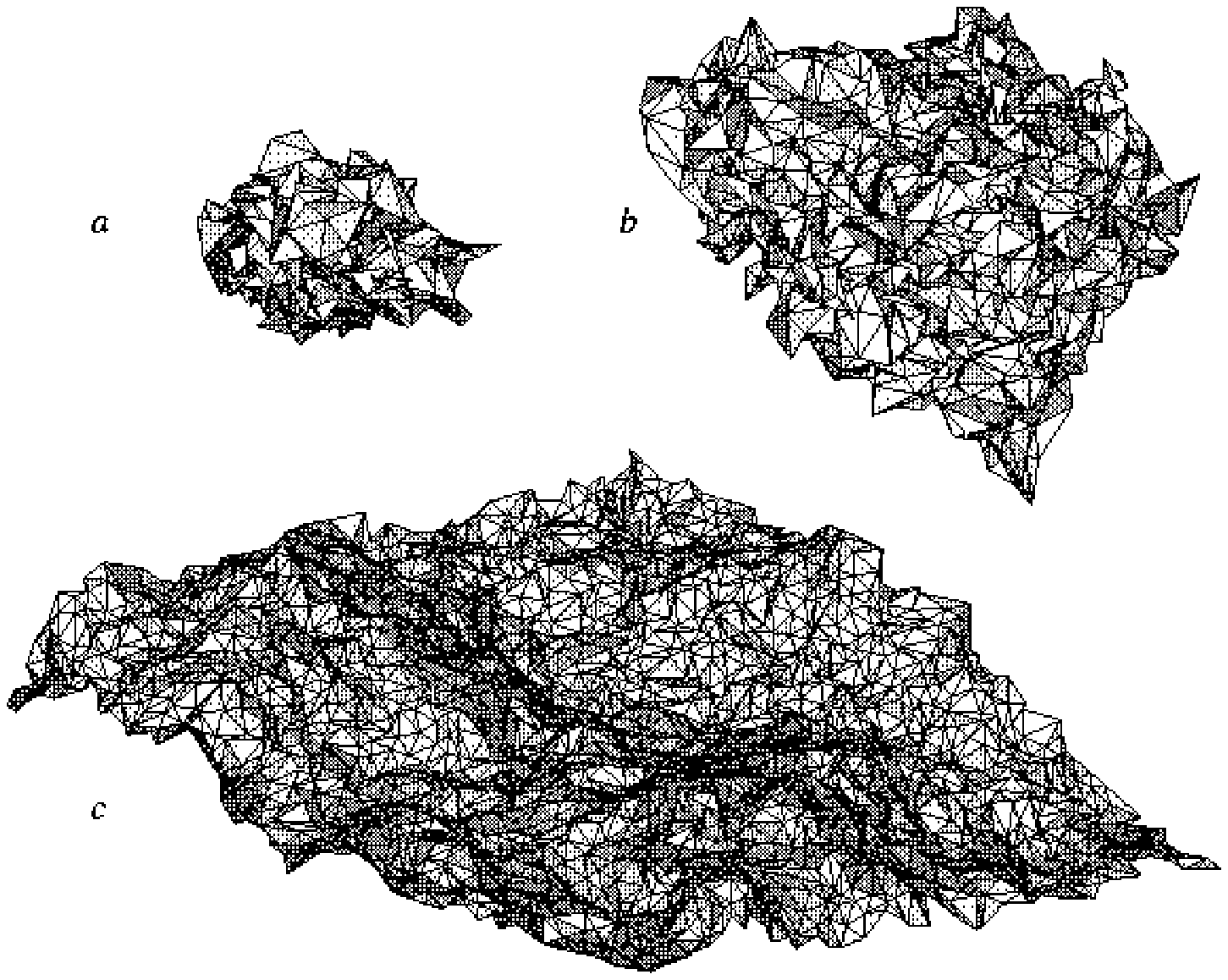}}
\protect\caption{Snapshot of the surfaces with $L = 46$ at {\it
(a)} $\kappa = 0.5$, {\it (b)} $\kappa = 0.8$ and {\it (c)} $\kappa =
2.0$. The average link length in each surface is comparable.}
\label{fig:confs}
\end{figure}
The Hausdorff dimension $d_H$ is related to the conventional Flory
exponent $\nu$ ($R_g \sim L^{\nu}$) via $d_H = 2/\nu$.  For a flat
surface, $\kappa > \kappa_c$, the radius of gyration scales linearly
with the internal size, and hence $d_H = 2$.  In the high-temperature
phase, on the other hand, $R_g$ scales logarithmically with the volume
of the surface, $R_g \sim \log(N)$.  In this case we say that the
Hausdorff dimension is infinite ($\nu = 0$). This justifies the
terminology {\em crumpled} phase. This behavior can be computed
exactly for $\kappa =0$ \cite{ADJ}, while mean field theory or
numerical methods are necessary for $\kappa < \kappa_c$
\cite{BEW,HW,WS,BET,ET1,ET2,Wheater,KR}. Experimentally one can
determine the Hausdorff dimension by measuring the structure function
of diffracted light.  A comparison with numerical simulations can be
found in Ref.\ \cite{AG}.

At the transition itself one might expect an intermediate behaviour
(semi-crumpled surfaces) with $d_H > 2$.  Indeed this has been
observed in \cite{ADJ}, where it is claimed that $d_H =4$ at
$\kappa_c$.

In Fig.\ \ref{fig:rg} we show our measurements of the radius of
gyration versus the bending rigidity for surfaces up to $64^2$ nodes.
As expected, we see a dramatic change in their spatial extent as we
pass through the phase transition.  The surfaces literally blow up.
This is better illustrated in Fig.\ \ref{fig:confs}, where we show
snapshots of the surfaces: {\it a} in the crumpled phase, {\it b} at
the phase transition and {\it c} in the flat phase.

We have only determined directly the Hausdorff dimension in the flat
phase (for $\kappa = 1.1$), where we have good statistics for surfaces
up to $128^2$ nodes.  The corresponding scaling plot is included in
Fig.\ \ref{fig:rg}.  A linear fit for surfaces larger than $16^2$
yields $d_H = 2.1(1)$, or $\nu = 0.95(5)$, as expected for a flat
surface.
\subsection{Normal-Normal Correlation Functions}
\label{sec:nncor}

As mentioned in Section \ref{sec:intro}, the different phases of a
crystalline membrane can also be distinguished by the behaviour of the
surface normals.  In the flat phase the normals will have true
long-range order, with the correlation function approaching a non-zero
asymptote.  In the crumpled phase the normals eventually decorrelate.

We define the normal-normal correlation function $G(r)$ as the
scalar product of two normals on the surface separated by
a distance of $r$ along the surface:
\begin{equation}
\label{eqn:corrdef}
G(r) \;=\; \langle {\bf n}_o \cdot {\bf n}_r \rangle.
\end{equation}
Here $o$ refers to the centre of the surface.
\begin{figure}
\centerline{\epsfxsize= \floatwidth  \epsfbox{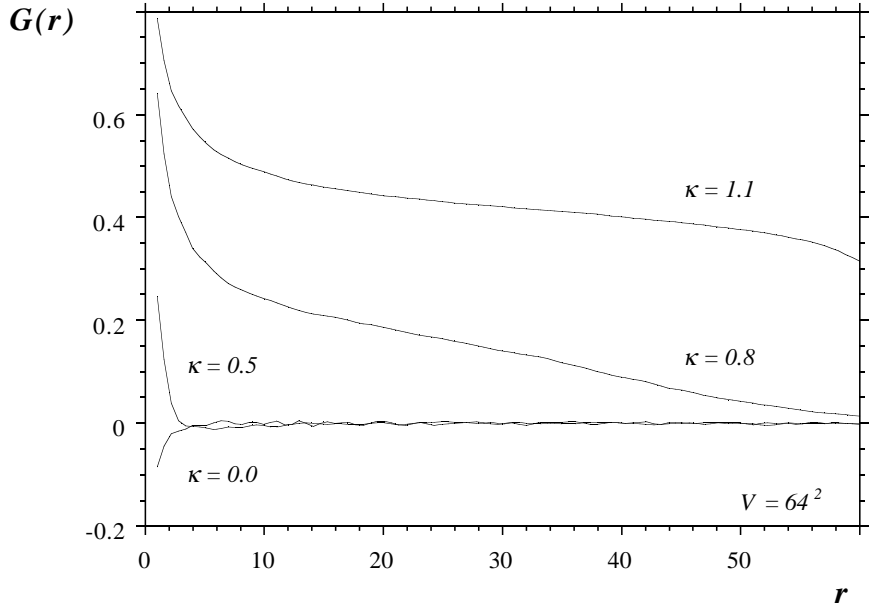}}
\caption{ The correlation function $G(r)$ for various values of the
bending rigidity $\kappa$.}
\label{fig:gvsk}
\end{figure}
In the crumpled phase we expect $G(r)$ to decay rapidly to zero. In
fact in the exactly solvable $\kappa=0$ Gaussian model one finds that
\begin{equation}
\label{eqn:crumpcorr}
G(r) = c_1 {\delta(r) \over r^2} - {c_2 \over r^4},
\end{equation}
where $c_1$ and $c_2$ are cutoff-dependent constants and $\delta(r)$
is the two-dimensional regularised $\delta$-function \cite{ADJ}.  In
the discrete case, the normals become decorrelated over a few lattice
steps. 

We have measured $G(r)$ for several bending rigidities, both in the
crumpled and flat phase.  In Fig.\ \ref{fig:gvsk} we illustrate this
on lattices with $64^2$ nodes.  For a more detailed discussion of how
we perform the measurements we refer to Appendix B.

In the crumpled phase $G(r)$ decays very rapidly to zero and, for
$\kappa = 0$, we indeed see a negative correlation between the normals
at short distances, as expected from Eq.\ (\ref{eqn:crumpcorr}).  On a
highly crumpled surface neighbouring triangles fold on each other.  As
$\kappa$ increases the normals become positively correlated at short
distance, although $G(r)$ still becomes negative (for $\kappa <
\kappa_c$) before decaying to zero.

At the critical point the normal-normal correlation function 
is expected to decay algebraically,
\begin{equation}
\label{eqn:critcorr}
G(r) \sim {1 \over r^{\bar\eta}},
\end{equation}
with an exponent $\bar\eta$ different from the Gaussian value 4.  This
is consistent with our numerical results.  A simple scaling argument
\cite{KN2} shows that $\bar\eta$ is related to the size (Flory)
exponent $\nu$ by $\bar\eta=4(1-\nu)$.  As we enter the critical
region ($\kappa \approx 0.79$) the normal-normal correlation function
still decays to zero, but now stays positive for all values of $r$.
It is also clear that it decays to zero more slowly than in the
crumpled phase.  A crude estimate of the exponent $\bar\eta$, using a
surface with $64^2$ nodes and $\kappa = 0.79$, yields $\bar\eta
\approx 0.71(5)$.  This implies a size exponent $\nu=0.82(1)$ and
hence Hausdorff dimension $d_H=2.4(2)$.  This is consistent with
previous numerical results \cite{KN2} and maybe be compared with two
theoretical predictions: $d_H=3$ from a $1/d$ renormalisation group
calculation \cite{DG}, applied to $d=3$, and $d_H=2.73$ from the
self-consistent screening approximation \cite{LDR}.

Finally, we show in Fig.\ \ref{fig:gvsk} a measurement of $G(r)$ for
$\kappa = 1.1$, i.e.\ in the flat phase.  In that case we see a
non-zero asymptote indicating true long-range-order in the normals.
Fitting $G(r)$ to an algebraic decay plus a constant term excludes
convincingly the possibility of a slow fall-off to zero, although
eventually $G(r)$ becomes zero due to boundary effects.  This
behaviour is found consistently for surfaces of various sizes.  We
will return to the behaviour of the normal-normal correlation function
in the flat phase in Section \ref{sec:flatcorr}.

\section{Flat Phase}
\label{sec:flat}
Here we shall describe the current theoretical understanding of the
flat phase of crystalline membranes. Given the Mermin-Wagner theorem
it is important to understand what could stabilise an ordered phase
in this two-dimensional system.  This is most easily described in the
Monge representation of a surface
\begin{equation}
\label{eqn:Monge}
{\bf x}_{\sigma} = h_{\sigma}\hat{z} + {\bf r}_{\sigma},
\end{equation}
where $h_{\sigma}$ is the height of the surface w.r.t.\ the base plane
$\hat{x}-\hat{y}$ and ${\bf r_{\sigma}}$ is the projection of ${\bf
x_{\sigma}}$ on the base plane.  We define the phonon fluctuations
${\bf u}_\sigma$ of the surface by
\begin{equation}
\label{eqn:phonons}
{\bf r}_\sigma = {\bf s}_\sigma + {\bf u}_\sigma
\end{equation}
where ${\bf s}_\sigma$ are the equilibrium positions of the nodes.
An effective Hamiltonian for the flat phase \cite{NP1,AN3} is a sum of
bending and elastic stretching energies
\begin{equation}
\label{eqn:contFE}
{\cal H}(h, {\bf u}) = \frac {\kappa}{2} \int d^2\sigma \,
{(\nabla^2h)}^2 + \frac {1}{2} \int d^2\sigma ( 2 \mu \, u_{\alpha
\beta}^2 + \lambda \, u_{\gamma \gamma}^2),
\end{equation}
where $\mu$ and $\lambda$ are the bare in-plane Lam\'e coefficients,
$\kappa$ is the bending rigidity and $u_{\alpha \beta}$ is the strain
tensor. This tensor measures the deformation of the induced
metric from the flat metric, and is given by
\begin{eqnarray}
\label{eqn:strain}
u_{\alpha \beta} &=& {1 \over 2} \left( \partial_\alpha {\bf x} \cdot
\partial_\beta {\bf x} - \delta_{\alpha \beta} \right) \\ & = & \frac
{1}{2} \{\nabla_\alpha u_\beta + \nabla_\beta u_\alpha +
\nabla_{\negthinspace \alpha}h\nabla_{\negthinspace \beta}h\}.
\end{eqnarray}
to linear order in {\bf u}.  The indices $\alpha$, $\beta$ run
over the internal coordinate {$\sigma = (\sigma_1,\sigma_2)$}. 

One can integrate out the phonon degrees of freedom by performing the
Gaussian integration\cite{jer2}.  One finds that the phonons give rise
to an effective long-range two-point interaction for the scalar
curvature.  This interaction flattens the surface by stiffening the
bending rigidity at large distances.

A more precise understanding may be obtained from self-consistent
calculations of the renormalised bending rigidity \cite{NP1,LDR},
mean-field calculations \cite{PKN}, the $\epsilon$-expansion
($\epsilon=4-D$, with $D$ the dimensionality of the surface)
\cite{AL,GDLP} and the $1/d$ expansion (where $d$ is the embedding
dimension) \cite{DG,PK}. Generally one finds that the renormalised
bending rigidity is scale dependent with a non-vanishing anomalous
dimension
\begin{equation}
\label{eqn:kappascales}
\kappa_R(q) \sim q^{-\eta}.
\end{equation}
At long wavelength $\kappa_R(q)$ is stiffened by short wavelength
undulations of the membrane.  It is also found that the elastic moduli
$\mu$ and $\lambda$ are softened at long wavelength
\begin{equation}
\label{eqn:lamescales}
\mu_R(q) \sim \lambda_R(q) \sim q^{\eta_u}.
\end{equation}
The exponents $\eta$ and $\eta_u$ are not independent. Rotational
invariance \cite{AL}, or self consistent integral equations for the
renormalised bending rigidity, imply
\begin{equation}
\label{eqn:etaetau}
\eta = 1 - \frac{\eta_u}{2}.
\end{equation}
An immediate implication of these anomalous dimensions or power-law
singularities are non-trivial roughness exponents and height and
in-plane displacement fluctuations or correlations. Defining a
roughness exponent $\zeta$ by the finite-size-scaling of the
mean-square height fluctuations in a box of size $L$
\begin{equation}
\label{eqn:zetadef}
\langle h^2 \rangle \; \sim \; L^{2\zeta},
\end{equation}
\noindent we see from Eq.\ (\ref{eqn:contFE}) that
\begin{equation}
\label{eqn:heta}
\langle h^2 \rangle \; \sim \; \int^{1\over a}_{1\over L}
\frac{d^2q}{(2\pi)^2} \frac{1} {\kappa_R(q) \, q^4} \; \sim \;
\int^{1\over a}_{1\over L} \frac{d^2q}{(2\pi)^2} \frac{1}
{q^{4-\eta}} \; \sim \; L^{2 - \eta},
\end{equation}
\noindent implying $\zeta = 1 - \eta/2$.  In the above $a$ is a
short-distance regularisation cut-off.  Similarly there is a
non-trivial exponent for phonon fluctuations \cite{AN2}
\begin{equation}
\label{eqn:omegadef}
\langle |{\bf u}^2| \rangle \; \sim \; 
L^{\eta_u}.
\end{equation}
In the framework of the self-consistent screening approximation it is
possible to obtain analytic predictions for these exponents
\cite{LDR}.  By assuming the scaling relations of Eqs.\
(\ref{eqn:kappascales}) and \ (\ref{eqn:lamescales}) one can sum the
terms in the perturbative expansion which renormalise $\kappa_R(q)$
and solve for the exponent $\eta$.  In subsequent sections we shall
describe our numerical measurements of these exponents.

Finally, a measure of long-range order in the flat phase is provided
by the normal-normal correlation function. In the Monge representation
a normal to the surface at point $\sigma$ is given by:
\begin{equation}
\label{eqn:normal}
{\bf n}_\sigma = \frac{\left( -\partial_{\sigma_1} h,
-\partial_{\sigma_2} h, \; 1 \right)} {\sqrt{1 + \vert \nabla h
\vert^2}},
\end{equation}
where $\sigma_1$ and $\sigma_2$ are the components of the intrinsic
coordinate $\sigma$.  To compute $\langle {\bf n}_\sigma \cdot {\bf
n}_o \rangle$ we rotate the surface so that the normal at the origin
coincides with the $\hat z$ axis.  Hence
\begin{equation}
\label{eqn:nndelh}
{\bf n}_\sigma \cdot \left( 0,\,0,\,1 \right) \; = \;
\frac{1}{\sqrt{1 + \vert \nabla h_{\sigma} \vert^2}} \; \approx \; 1 -
\frac{1}{2} \vert \nabla h_{\sigma} \vert^2.
\end{equation}
The correlation function then follows from the height-field
propagator,
\begin{equation}
\label{eqn:nnfalloff}
\langle {\bf n}_\sigma \cdot {\bf n}_o\rangle \;  \simeq \; 
1\;-\;\frac{1}{2} \int_{1 \over r}^{1 \over a} \frac{{\rm
d}^2q}{(2\pi)^2} \; \frac{q^2}{\kappa_R(q)\;q^4}\;
 \approx \; C + \frac{c}{r^\eta}\, ,
\end{equation}
where $r$ is the geodesic distance between the point at $\sigma$ and
the origin $o$ in the embedding space.  Thus a power-law decay of the
normal-normal correlation function to a non-zero asymptote provides a
direct measure of the bending rigidity anomalous exponent $\eta$.  
\subsection{Shape Tensor}
\label{sec:inertia}
\begin{figure}
\centerline{ \epsfxsize=\floatwidth \epsfbox{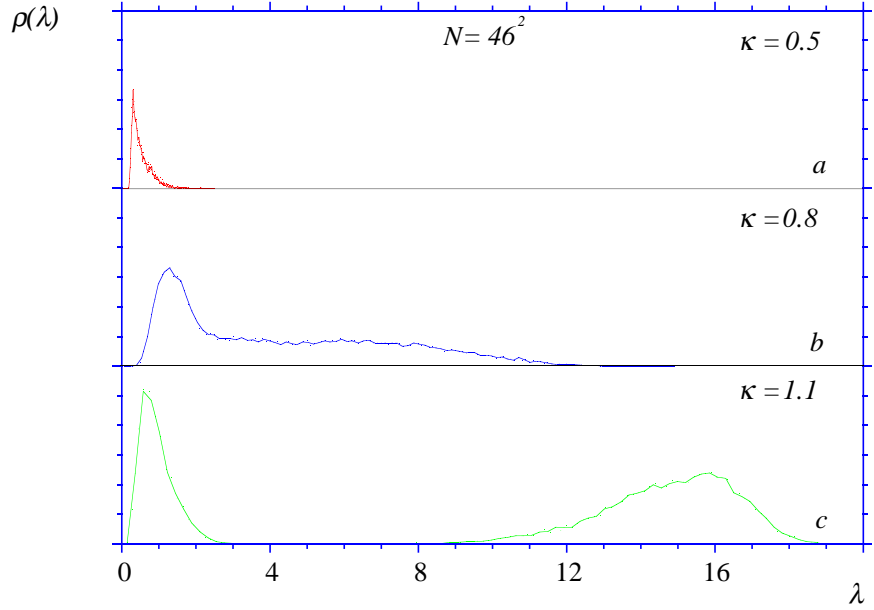}} \caption{The
distribution of eigenvalues $\rho(\lambda)$ of the inertia tensor.
The number of nodes is $46^2$ and the hexagon radius is 17.}
\protect\label{fig:eigs}
\end{figure}
We now return to our numerical results.  Detailed information about
the shape of the surface in the embedding space is provided by
examining averages of the eigenvalues of the inertia tensor. More
precisely we study the shape tensor, defined as the anisotropic part
of the inertia tensor:
\begin{equation}
\label{eqn:inertia}
I_{ij}[{\bf x}] = \frac{1}{3N} \sum_{\bf \sigma} x_{i}({\bf \sigma})
x_{j}({\bf \sigma}),
\end{equation}
where $i$ and $j$ refer to the components of ${\bf x}$.  The full
inertia tensor has an additional isotropic contribution proportional
to $\delta_{ij}$ --- this scales like $R_g^2$.  As the shape tensor
$I_{ij}$ is strongly influenced by the boundary (in fact contributions
from the boundary region dominate the sum Eq.\ (\ref{eqn:inertia})),
we restrict our measurements to a hexagonal subset of the mesh, as
described in Section \ref{sec:boundary}. If we refer the surface to
the body-fixed frame defined by the eigenvectors and its centre of
mass, the eigenvalues of $I_{ij}$, $\lambda_i$, are nothing but the
dispersion of the $i$-th component of ${\bf x}$, averaged over the
surface
\begin{equation} 
\label{eqn:eigs}
\lambda_i = \left\langle \frac{1}{3N_r} \sum_{\sigma \in H_r}
x^{2}_i(\sigma)  \right\rangle,
\end{equation}
where $N_r$ is the number of nodes inside a hexagon of radius $r$. 

We obtain the eigenvalues $\lambda_i$ by diagonalising $I_{ij}$.  The
distribution of these eigenvalues, $\rho(\lambda)$, is distinctly
different in the two phases.  In Fig.\ \ref{fig:eigs} we show three
examples of this.  In the crumpled phase ({\it a}) $\rho(\lambda)$ has
a single peak, implying a three-fold degeneracy of the
eigenvalues\footnote{In fact this degeneracy is not likely to be
exact.  In a body-fixed frame there is always a hierarchy of
eigenvalues.  Fig.~\ref{fig:eigs}{\it a} shows that this effect is
very small.}.  This is a simple consequence of the isotropy of a
crumpled surface in the embedding space.  The position of the peak
indicates the average extension of the surface, while its width is
related to the fluctuations about that value.  At the phase transition
({\it b}) there is still a single peak in $\rho(\lambda)$, but now
accompanied by a significant tail extending to large eigenvalues.
This is due to increasing fluctuations in the size of the surface in
the critical region.

The behaviour is very different for a flat surface ({\it c}).  We now
see two well resolved peaks in $\rho(\lambda)$ indicating, as
expected, a broken $O(3)$ symmetry. There is a single minimum
eigenvalue, corresponding to the left peak, which can be identified
with the average square thickness of the surface.  But there are also
two almost degenerate large eigenvalues, as can be established by 
measuring
the area of the right hand peak.  This is a result of the remnant
$O(2)$ symmetry in the plane. If we did not restrict our
measurements to a hexagonal subset of the mesh, we would actually see
three peaks in $\rho(\lambda)$, due to the anisotropy of the
boundary.
\subsection{Roughness Exponent}
\begin{figure}
\centerline{ \epsfxsize= \floatwidth \epsfbox{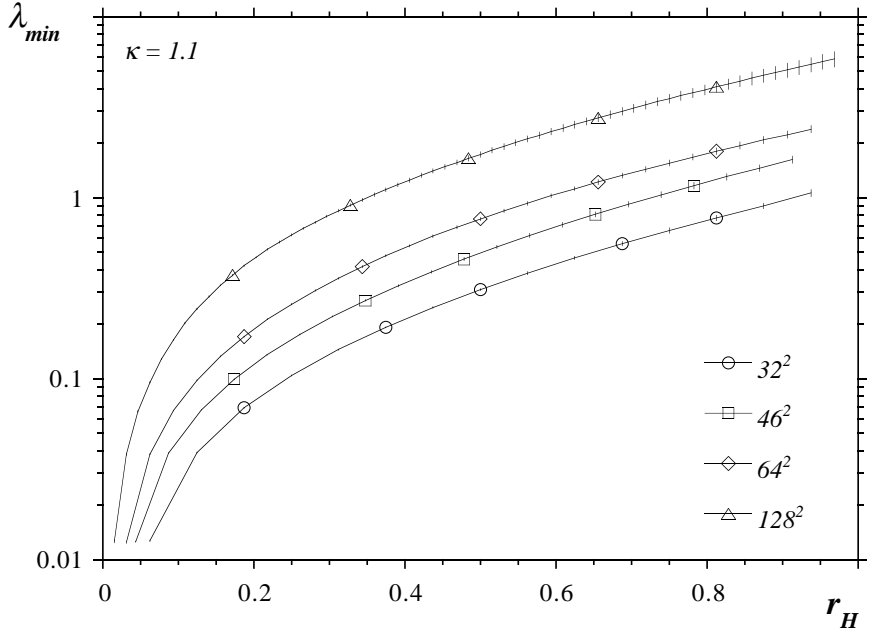}}
\protect\caption{The minimum eigenvalue $\lambda_{min}$ versus the
radius of the hexagon for various lattice sizes and $\kappa = 1.1$.
The solid lines are obtained from a polynomial fit and the points are
simply to guide the eye.}  \protect\label{fig:rough}
\end{figure}
To measure the height fluctuations in the flat phase, and the
corresponding roughness exponent $\zeta$, we need an estimate of the
out-of-plane fluctuations.  This is provided by the minimum eigenvalue
of the shape tensor which, as discussed in last section, is simply
the average squared height.  Hence
\begin{equation}
\label{eqn:zetaaa}
\lambda_{min} \; \sim \; \left\langle \int d^2 {\sigma} \, h^2_\sigma
\right\rangle \; \sim \; L^{2\zeta}.
\end{equation}
In Fig.\ \ref{fig:rough} we plot the minimum eigenvalue
$\lambda_{min}$ versus the (normalised) radius of the hexagonal
subsets we used, $r_H = 2r/L$.  This is shown for four different size
surfaces at $\kappa = 1.1$. A comment on how we treat the data: as we
can only use hexagonal subsets of certain radii (discrete units of the
lattice spacing), measurements on surfaces of different sizes will
yield measurements at different values of the normalised radius $r_H$.
To compare measurements from surfaces of different sizes, at the
{\it same} value of $r_H$, we must therefore interpolate between the
data points.  These are the solid lines in Fig.\ \ref{fig:rough}.  For
the interpolation we used a polynomial fit, with the degree of the
polynomials sufficiently large to yield a stable fit.  We checked that
different interpolation methods did not affect the results.
\begin{figure}
\epsfxsize= \floatwidth \centerline{ \epsfbox{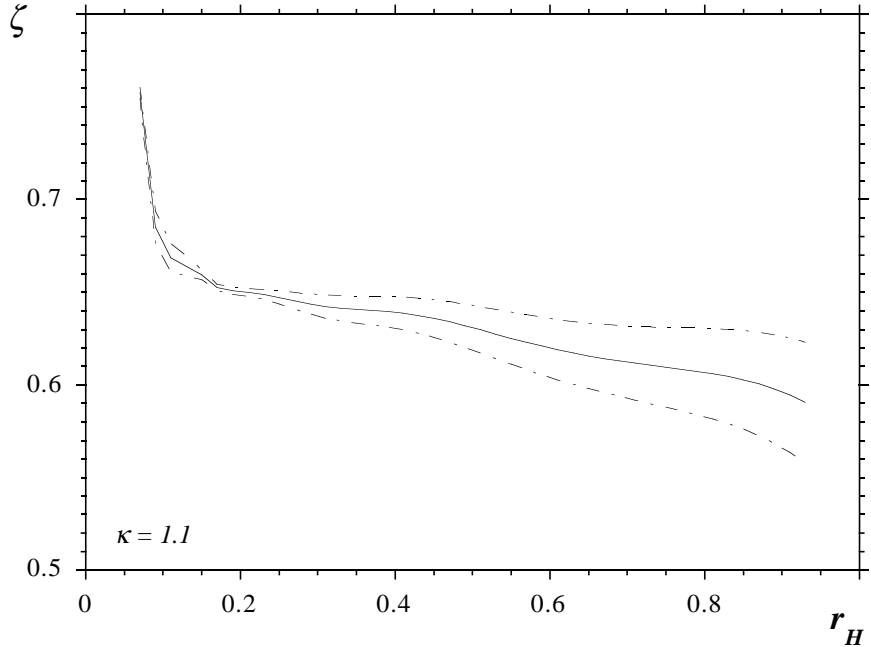}}
\caption{The roughness exponent $\zeta$ versus the normalised radius
of the hexagon.  We extract the value of $\zeta$ from the plateau
region.  The dashed lines indicate the size of the error bars. The
bending rigidity is $\kappa = 1.1$.}  \protect\label{fig:zeta}
\end{figure}
The roughness exponent $\zeta$ is then determined from the finite-size
scaling of the minimum eigenvalue for a fixed value of $r_H$.  The
result is shown in Fig.\ \ref{fig:zeta}. The solid line is
$\zeta(r_H)$ while the dashed lines indicate the error.  There are
large discretisation effects for small values of $r_H$, as expected.
For $0.2 \leq r_H \leq 0.5$ there is a reasonably stable value of the
roughness exponent.  Our best estimate from this intermediate region
is $\zeta = 0.64(2)$. This should be compared to the theoretical
predictions $\zeta = .590$ \cite{LDR} and $\zeta = 2/3$ \cite{DG} and
from measurements on the spectrin network $\zeta = 0.65(10)$
\cite{Skel}. Previous numerical investigations of models of tethered
surfaces have found a wide range of values for $\zeta$.  These include
0.5 \cite{LG}, 0.53 \cite{A}, 0.56(2) \cite{Skel}, 0.6 \cite{ZDK}, 0.64
\cite{AN1,ARP,LM}, 0.65 \cite{AN2,AK} and 0.70 \cite{GK,KG}.  From the
scaling relation $\eta = 2(1-\zeta)$ we find $\eta=0.72(4)$. For
larger values of $r_H$ we see clear evidence of boundary effects.
Indeed, if we did not work with hexagonal subsets, we would not be
able to extract a reliable estimate for $\zeta$.

We have also measured $\zeta(r_H)$ for $\kappa = 2.0$, although the
largest surface simulated in this case had only $64^2$ nodes. Once
again a finite-size-scaling analysis yields a stable value of $\zeta$
in the same interval of $r_H$.  In this case the result is $\zeta =
0.71(2)$. This value is 3$\sigma$ larger than the corresponding value
at the same lattice size for $\kappa = 1.1$. We believe that this is
due to larger finite size effects and that our results are consistent
with having universal critical exponents in the flat phase as
predicted by the fixed point of Refs.\ \cite{AL,AGL,GDLP,GDLP2}.
\subsection{Phonon Fluctuations}
\label{sec:phonon}
\begin{figure}
\centerline{\epsfxsize=\floatwidth \epsfbox{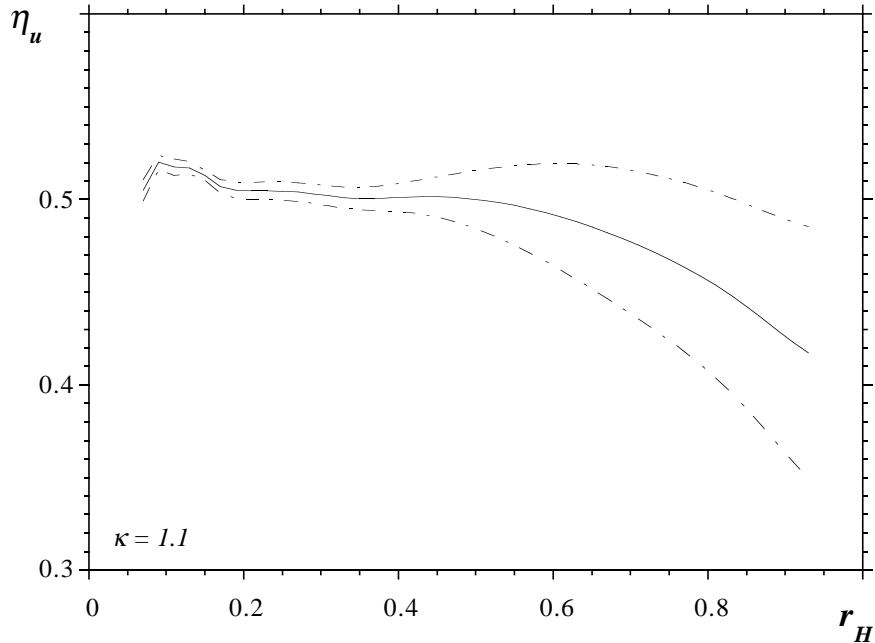}} 
\protect\caption{The
exponent $\eta_u$ as a function of the hexagon radius. The dashed lines
indicate the size of the error bars.}
\protect\label{fig:omega}
\end{figure}
We now examine the issue of the in-plane elastic degrees of freedom in
the flat phase of our model. For convenience, we rotate the surface so
that the eigenvector associated to the smallest eigenvalue points in
the $\hat{z}$ direction.  Projecting the surface onto the
$\hat{x}{-}\hat{y}$ plane gives us a discretised analogue of the field
${\bf r}_{\sigma}$ of Eq.\ (\ref{eqn:Monge}).

The first step in the analysis must be to determine the phonon field
{\bf u}$_\sigma$. We therefore need to determine the field ${\bf
s}_\sigma$ giving the equilibrium positions of the nodes.  As before,
we restrict our analysis to the hexagonal subsets $H_r$ of Section
\ref{sec:boundary}. We assume that the equilibrium positions of the
nodes lie exactly on a regular hexagon in the $\hat{x}{-}\hat{y}$
plane. In the course of the Monte Carlo simulation, the surface
fluctuates in the embedding space so that the orientation of its
principal axes and its overall extent change constantly. Thus we need
to find the regular hexagon ${\bf s}_{\sigma}$ for each configuration
we analyse.  We define the functional
\begin{equation}
\label{eqn:functional}
F \; = \; \sum_{\sigma} \left({\bf r}_\sigma - {\bf s}_\sigma\right)^2
\; = \; \sum_{\sigma} {\bf u}_\sigma^2, 
\end{equation}
and we choose the equilibrium position of the mesh to be best
represented by the hexagon which minimises $F$.  The regular hexagon
{\bf s} can be parametrised by the position of its centre, a rotation
angle $\theta$ and a scale factor $\xi$.  The centre of the hexagon is
trivially set to coincide with the centre of mass of the projected
surface.  Minimizing the functional $F$ with respect to the angle
$\theta$ and the scale $\xi$ yields a unique solution up to the
six-fold symmetry of the regular hexagon $\theta \to \theta + \pi/6$.
This six-fold degeneracy of the minimum of $F$ can be eliminated by
requiring that the internal labels of the ${\bf s_\sigma}$ overlap
with the ones of the ${\bf r}_\sigma$.

We determine the characteristic scaling exponent $\eta_u$, defined in
Eq.\ \ref{eqn:omegadef}, by a power law fit of $\langle {\bf u}^2
\rangle_{r_H}$ to $L$, for fixed values of $r_H$.  Fig.\
\ref{fig:omega} shows the result of the fit for $\kappa = 1.1$.  We
extract the value of the exponent from the plateau in the figure. This
gives $\eta_u = 0.50(1)$ at $\kappa = 1.1$.  The corresponding value
of $\eta$ obtained from the scaling relation of Eq.\
(\ref{eqn:etaetau}) is $\eta = 0.750(5)$.  We also quote the results
of the same analysis at $\kappa = 2.0$: $\eta_u = 0.40(1)$.
\subsection{Correlations in the Flat Phase}
\label{sec:flatcorr}
\begin{figure}
\centerline{ \epsfxsize=\floatwidth \epsfbox{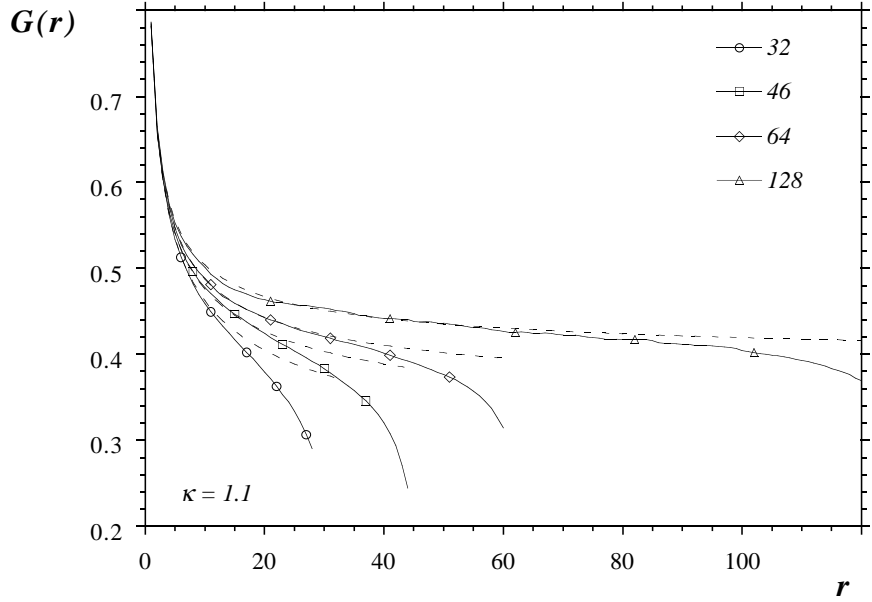}}
\caption{The $G(r)$ fall-off for $L=32$, $46$, $64$ and
$128$. Representative data points are shown to guide the eye. The
dashed lines are the best fits to Eq.\ (\ref{eqn:nnfalloff}).}
\label{fig:falloff}
\end{figure}
\begin{figure}
\centerline {\epsfxsize=\floatwidth \epsfbox{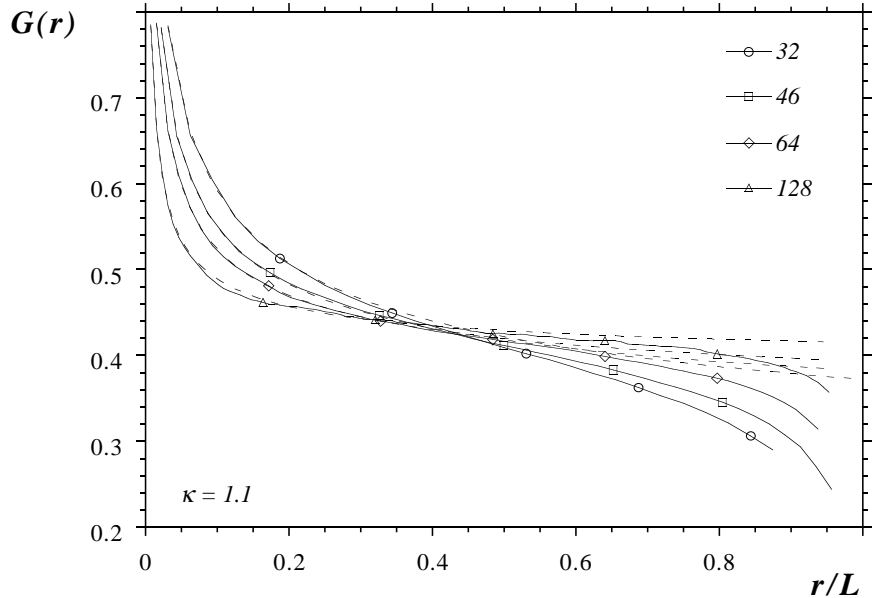}} \caption{A
plot of $G(r)$ with $r$ rescaled to show the finite-size effects on 
the
asymptote ($\kappa = 1.1$).  This rescaling demonstrates that the 
asymptote
stabilizes as the lattice size increases.} 
\protect\label{fig:asymp}
\end{figure}

In this section we treat the normal-normal correlation function in the
flat phase. The measured correlation functions are shown in Fig.\
\ref{fig:falloff}. We fit this correlation function to the expected
power-law behaviour of Eq.\ (\ref{eqn:nnfalloff})
\begin{equation}
\label{eqn:fit}
G(r) = C + \frac{B}{r^\eta} ,
\end{equation}
using a correlated least squares algorithm. We find a non-zero
asymptote $C$ whose value tends to increase with lattice size. This is
shown in Fig.\ \ref{fig:asymp}.
The correlated least squares algorithm attempts to minimise a
chi-squared function of the form
\begin{equation}
\label{eqn:chisq}
\chi^2=\frac{1}{N(N-P)}\sum_{ij}^N (y_i-y^*_i(p_i)
)C_{ij}(y_j-y^*_j(p_i)).
\end{equation}
The data is fitted to a functional form $y^*_i(p_i)$ where
$i$ is the (lattice) geodesic distance and the quantities $\{p_i\},
i=1\ldots P$, are the variational parameters in the fit. The matrix
$C_{ij}$ is the inverse of the correlation matrix
\begin{equation}
C_{ij}=\left(M^{-1}\right)_{ij}
\end{equation}
where
\begin{equation}
M_{r r^\prime} = \left\langle
G\left(r\right)\,G\left(r^\prime\right)\right\rangle_c.
\end{equation}
Note that for uncorrelated data the matrix $M$ is diagonal, and Eq.\
(\ref{eqn:chisq}) reduces to the usual definiition of $\chi^2$.  The
inversion of this matrix is typically a delicate operation.  Because
of limited statistics it will often be close to singular and the
smallest eigenvalues will only be poorly estimated from the data. In
this situation it is necessary, and indeed correct, to restrict the
inversion to an appropriate subspace that is spanned by eigenvectors
with eigenvalues which are large enough to be estimated reliably from
the data. This can be achieved through singular value decomposition
techniques. The dimension of the subspace is referred to as the
singular value decomposition (SVD) cut. For more details we refer the
reader to \cite{CDDH}.

In assessing the results of the fitting procedure we examined cases
with a range of values of initial and final distances and SVD cuts.
To obtain a $\chi^2$ of order unity we usually had to discard data
from the last third of the path to the boundary and roughly one
quarter of the eigenvalues.
\begin{table}
\begin{center}
\begin{tabular}{|l|r@{.}l r@{.}l r@{.}l |r@{.}l r@{.}l r@{.}l|}\hline
&\multicolumn{6}{c|}{$\kappa$ = 1.1}&\multicolumn{6}{c|}{$\kappa$ =
2.0}\\ $L$ & \multicolumn{2}{c}{$C$} & \multicolumn{2}{c}{$\eta$} &
\multicolumn{2}{c|}{$\chi^2/${\it dof}} & \multicolumn{2}{c}{$C$} &
\multicolumn{2}{c}{$\eta$} & \multicolumn{2}{c|}{$\chi^2/${\it dof}}\\
\hline 
32 & 0&181(16) & 0&331(7) & 8&6 & 0&240(32) & 0&141(5) & 14&8\\ 
46 & 0&274(9) & 0&397(5) & 1&66 & 0&309(21) & 0&154(4) & 2&32\\ 
64 & 0&321(5) & 0&447(6) & 0&96 & 0&448(30) & 0&203(4) & 2&19\\ 
128 & 0&383(6) & 0&521(6) & 1&14 &\multicolumn{6}{c|}{---}\\\hline
\end{tabular}
\caption{Fit to the correlation function $G(r)$.}
\label{tbl:fitG}
\vskip 3em
\begin{tabular}{|l|r@{.}l r@{.}l r@{.}l |r@{.}l r@{.}l r@{.}l|}\hline
& \multicolumn{6}{c|}{The constant $C$} & \multicolumn{6}{c|}{The
exponent $\eta$}\\ $r_{min}$ & \multicolumn{2}{c}{$L = 46$} &
\multicolumn{2}{c}{$L = 64$} & \multicolumn{2}{c|}{$L = 128$} &
\multicolumn{2}{c}{$L = 46$} & \multicolumn{2}{c}{$L = 64$} &
\multicolumn{2}{c|}{$L = 128$}\\ \hline 
1 & 0&274(9) & 0&321(5) & 0&383(6) & 0&397(5) & 0&458(7)& 0&521(6)\\ 
2 & 0&262(15) & 0&325(7) & 0&388(7) & 0&376(10) & 0&492(9) & 
0&546(9)\\ 
3 & 0&262(17) & 0&327(7) & 0&405(7) & 0&388(14) & 0&481(9) & 
0&658(15)\\\hline
\end{tabular}
\end{center}
\caption{Effect of the short distance data ($\kappa = 1.1$).}
\label{tbl:short}
\end{table}

In Table \ref{tbl:fitG} we show the fitted values of the asymptote $C$
and the exponent $\eta$. The fits are obtained including all the short
distance data and we show only errors from the fit.  It is clear that
a more important source of error is finite size effects.  Clearly both
the constant $C$ and the exponent $\eta$ exhibit a shift with $L$ well
in excess of the fitted errors.  But this shift is systematic and
indeed can be quantified by assuming a naive $1/L$ correction term to
the correlation function $G(r)$.  In that case the leading correction
to $C$ will be $1/L$, while it is $\log(L)/L$ for the exponent $\eta$.
This conjecture is in excellent agreement with our data, at least for
$\kappa = 1.1$, and implies infinite volume values of $C \approx 0.45$
and $\eta \approx 0.62$.  This is in quantitative agreement with our
previous estimates.

Another source of systematic error stems from discarding some of the
short distance data.  In Table \ref{tbl:short} we show the results of
fits to C and $\eta$ obtained by discarding data at distance $r <
r_{min}$.  It is clear that the values change with $r_{min}$, although
not a great deal. Discarding short distance data increases the value
of $\eta$ for the large lattices.  This improves the comparison with
the expected value.

Finally, a few comments on the data at $\kappa = 2.0$.  The values of
$\eta$ quoted in Table \ref{tbl:short} do not agree well with the
values obtained at $\kappa = 1.1$.  But it should be mentioned that
the fits to Eq.\ (\ref{eqn:fit}) are not as good in this case, which
is reflected in higher values of the $\chi^2$.

\section{Conclusions}
In this paper we used a large scale Monte Carlo simulation to show
that an extremely simple model of a crystalline membrane has a
well-defined flat phase.  In this phase the critical exponents are
consistent both with previous simulations of tethered membranes and
with theoretical predictions. In particular, the model we study has
dynamically generated elastic moduli. The flat phase is convincingly
demonstrated by the existence of a non-zero asymptote for the
normal-normal correlation function, which strengthens with increasing
system size. The flat phase exponents we find at $\kappa = 1.1$ are:
size (Flory) exponent $\nu = 0.95(5)$ ($d_H = 2.1(1)$), roughness
exponent $\zeta = 0.64(2)$, $\eta_u = 0.50(1)$. A check on the
consistency of our results for $\zeta$ and $\eta_u$ is obtained by
comparing their respective scaling predictions for $\eta$.  Our value
of $\zeta$ implies $\eta\approx 0.72$ and our value of $\eta_u$
implies $\eta\approx 0.75$. These are consistent within our
statistical errors. A direct measurement of $\eta$ from the power law
decay of the normal-normal correlation functions is more difficult and
is discussed in Section \ref{sec:flatcorr}. For $L = 128$ and $r_{min}
= 3$ we obtain $\eta = 0.658(15)$. At higher bending rigidity we
obtain somewhat different exponents, but we believe this is due to
larger finite-size effects.

For completeness, we also establish that the model we consider has a
crumpling transition. Preliminary determination of the critical
exponents at the transition gives results consistent with existing
simulations for related models.

\subsection*{Acknowledgments}
David Nelson has contributed to this work with many suggestions and
comments.  We would also like to thank Mehran Kardar, Emmanuel
Guitter, Alan Middleton, Paul Coddington and Gerard Jungman for
helpful discussions.  Enzo Marinari kindly provided the random-number
generator routine used in our parallel code.  We are grateful to NPAC
(North-East Parallel Architecture Center) for the use of their
computational facilities.  The research of MB and MF was supported by
the Department of Energy U.S.A.\ under contract No.\
DE-FG02-85ER40237.  SC and GT were supported by research funds from
Syracuse University.  Part of the work of KA was done at the Institute
for Fundamental Theory at Gainesville and was supported by DOE grant
No.\ DE-FG05-86ER-40272. We also acknowledge the use of the software
package Geomview for membrane visualisation and code development
\cite{Geomview}.

\section*{Appendix A: Elastic Constants in the Gaussian Model}
As we noted in the introduction, there are several potential
pathologies in a model in which the tethering potential contains no
hard core repulsion, such as the one we treat in this paper.  The
effective elastic constants may vanish or be too weak to generate a
stable flat phase.  Even if the model possesses a flat phase it may
belong to a different universality class from a model with bare
elastic constants.

The simplest argument supporting such concerns arises from a mean
field theory calculation of the elastic moduli of our discrete
tethering potential along the lines of \cite{SN}.  Consider an
equilibrium spring length parameter $a$ in the pair potential of Eq.\
(\ref{eqn:ouraction}):
\begin{eqnarray}
{\cal H}_T(a) &=& \sum_{\langle \sigma \sigma^\prime \rangle} {\left( 
\left|
{\bf x}_{\sigma} - {\bf x}_{\sigma^\prime}\right| - a\right)^2}\\ &=&
{1\over 2}\sum_{\sigma} \sum_{\sigma^\prime} \left( \left| {\bf 
x}_{\bf
\sigma} - {\bf x}_{\sigma^\prime}\right| - a\right)^2\, .
\label{eqn:kkkouraction}
\end{eqnarray}
\noindent
In the second sum $\sigma^\prime$ runs over the neighbour nodes of
$\sigma$.  Note that we have chosen units such that the spring
constant $k/(k_B T) = 1$ and {\bf x}, {\bf u} {\em and} ${\bf \sigma}$ are
dimensionless.  For $a$ sufficiently large we may describe the
location of the nodes ${\bf x}_{\bf \sigma}$ by
\begin{equation}
\label{eqn:kkkudef}
\left( {\bf x}_{\bf \sigma} - {\bf x}_{\rm {\sigma}'} \right)_\alpha =
\left(\delta_{\alpha\beta} + u_{\alpha\beta}\right) {\bf s}_{\bf
\sigma\sigma^\prime}^\beta ,
\end{equation}
where ${\bf s}_{\bf\sigma}$ spans a regular hexagonal lattice, 
${\bf s}_{\bf\sigma\sigma^\prime}= {\bf s}_{\bf\sigma}-{\bf
s}_{\bf\sigma^\prime}$ and $|{\bf s}_{\bf\sigma\sigma^\prime}|= a$.  
$\alpha$ and
$\beta = 1,2$ are intrinsic coordinate indices on the surface. Then
\begin{eqnarray}
\label{eqn:kkkxno}
\left|{\bf x}_{\bf \sigma} - {\bf x}_{\rm {\sigma}'}\right|^2 & = &
\left(\delta_{\alpha\beta}
+2u_{\alpha\beta}+u_{\alpha\gamma}u_{\beta\gamma}\right) {\bf
s}_{\bf\sigma\sigma^\prime}^\alpha {\bf
s}_{\bf\sigma\sigma^\prime}^\beta\\ 
& = &
a^2 +\left(2u_{\alpha\beta}+u_{\alpha\gamma}u_{\beta\gamma}\right) 
{\bf
s}_{\bf\sigma\sigma^\prime}^\alpha {\bf s}_{\bf\sigma\sigma^\prime}^\beta.
\end{eqnarray}
At this point we redefine $u'_{\alpha\beta} = u_{\alpha\beta} + {1
\over 2} u_{\alpha\gamma}u_{\beta\gamma}$ and subsequently drop the
primes. This will not affect the end result to quadratic order. 
Expanding for small fluctuations $u_{\alpha\beta} \ll 1$ we have
\begin{eqnarray}
\label{eqn:kkkxnom}
\left|{\bf x}_{\bf \sigma} - {\bf x}_{\rm {\sigma}'}\right| & = & 
\left(a^2
+ 2 u_{\alpha\beta} {\bf s}_{\bf\sigma\sigma^\prime}^\alpha {\bf
s}_{\bf\sigma\sigma^\prime}^\beta\right)^{1\over 2}\\ & \approx & a 
+{1\over a}
u_{\alpha\beta} {\bf s}_{\bf\sigma\sigma^\prime}^\alpha {\bf
s}_{\bf\sigma\sigma^\prime}^\beta - {1\over 2
a^3}u_{\alpha\beta}u_{\gamma\delta} {\bf s}_{\bf\sigma\sigma^\prime}^\alpha
{\bf s}_{\bf\sigma\sigma^\prime}^\beta {\bf s}_{\bf\sigma\sigma^\prime
}^\gamma
{\bf s}_{\bf\sigma\sigma^\prime}^\delta + \ldots.
\end{eqnarray}
To evaluate Eq.\ (\ref{eqn:kkkouraction}), consider the 6 unit
lattice vectors ${\bf d}_b$ of a regular hexagonal lattice where $b=
1,\ldots ,6$. Performing the sum over the neighbours 
${\bf\sigma^\prime}$
gives
\begin{eqnarray}
\label{eqn:kkkactc}
{\cal H}_T(a) & = & {1\over 2} \sum_{\bf\sigma} \sum_{b=1}^6 \left(a
 u_{\alpha\beta}{\bf d}_b^\alpha{\bf d}_b^\beta\right)^2\\ & = & 
{3\over 8}
 a^2 \sum_{\bf\sigma} \left(\delta^{\alpha\gamma}\delta^{\beta\delta 
} +
 \delta^{\alpha\delta}\delta^{\beta\gamma } + \delta^{\alpha\beta
 }\delta^{\gamma\delta} \right) u_{\alpha\beta} u_{\gamma\delta}\\ 
& = & {3\over 8} a^2 \sum_{\bf\sigma} \left( 2
 u_{\alpha\beta} u_{\beta\alpha} + u_{\gamma\gamma}^2\right).
\end{eqnarray}
For sufficiently large $N$, and $a$ fixed, one can approximate the
above discrete sum with an integral. 
\begin{equation}
\label{eqn:kkkcntlm}
{\cal H}_T(a) \approx \int \frac{d^2\xi}{\epsilon^2}
\left({\sqrt{3}\over2}a^2 u_{\alpha\beta} u_{\beta\alpha} +
{\sqrt{3}\over4}a^2 u_{\gamma\gamma}^2\right),
\end{equation}
where $\epsilon$ is the intrinsic lattice spacing.  Often one is
interested in the case $\epsilon \simeq a$.  We keep the $\epsilon$
and $a$ dependence since we are interested in the case $a = 0$,
$\epsilon \neq 0$.  

To determine the elastic constants recall that the most general continuum
action for a crystalline membrane is \cite{PKN}
\begin{eqnarray}
\label{eqn:kkkactcont}
\nonumber {\cal H} & = & \int d^2\xi\left[ {\kappa\over
2}\partial^2{\bf x}\cdot\partial^2{\bf x} + u\left( \partial_\alpha
{\bf x}\cdot\partial_\beta {\bf x}\right)^2 \right. + \\ & & + \left.
v\left( \partial_\gamma {\bf x}\cdot\partial_\gamma {\bf x}\right)^2
+ {t\over 2} \left( \partial_\alpha {\bf x}\cdot\partial_\beta {\bf
x}\right) \right] \\ & = & \int d^2\xi\left[ {\kappa\over
2}\partial^2{\bf x}\cdot\partial^2{\bf x} + \mu u_{\alpha\beta}
u_{\beta\alpha} + {\lambda\over 2} u_{\gamma\gamma}^2 + \tau
u_{\gamma\gamma} \right],
\end{eqnarray}
where $u_{\alpha\beta} = {1\over 2} (\partial_\alpha{\bf
x}\cdot\partial_\beta{\bf x} - \delta_{\alpha\beta})$ is the strain
tensor Eq.\ (\ref{eqn:strain}). The couplings in the two expressions
in Eq.\ (\ref{eqn:kkkactcont}) are related by
\begin{equation}
\label{eqn:kkkcoupr}
u = {\mu\over 4}\, , \qquad v = {\lambda\over 8} \, , \qquad t =
\tau - \mu - \lambda.
\end{equation}
Thus the bare elastic constants, dimensionless in our units, arising
from the lattice action (\ref{eqn:kkkouraction}) are
\begin{equation}
\label{eqn:kkkcoupd}
\mu = \lambda = {\sqrt{3}\over 4} \frac{a^2}{\epsilon^2}\, ,\qquad \tau=0.
\end{equation}
Notice that if $\epsilon = a$ the elastic constants are independent of
the lattice spacing.  The model we have studied corresponds to the
limit $a\rightarrow 0$ and $\epsilon$ fixed. In this limit
fluctuations become large and the above calculation breaks down.

\section*{Appendix B: Measuring Correlation Functions}
\label{sec:distance}
\begin{figure} 
\centerline{ \epsfxsize=\floatwidth \epsfbox{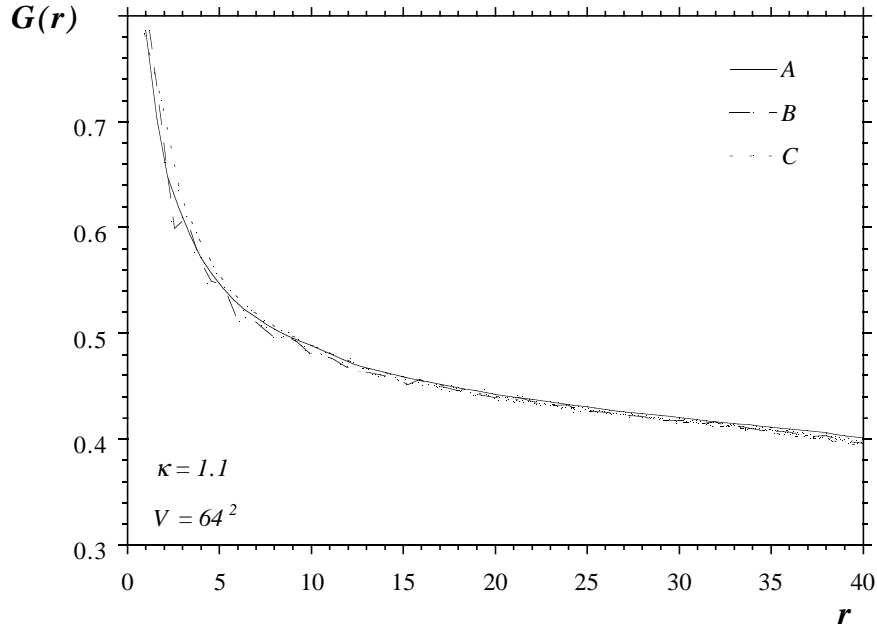}}
\caption{The comparison of the three methods used to measure the
geodesic distance $r$. Methods {\it A} and {\it B} use the intrinsic
metric. For method {\it A} we construct a piecewise linear path
through the centre of each triangle, while for method {\em B} we
construct a straight line between the centre of the starting and
ending triangle. Method {\em C} is the same as {\it A}, except that it
uses the induced metric to compute the distances between the
triangles.}
\label{fig:geodes}
\end{figure}
A necessary premise for measuring the normal-normal correlations
function $G(r)$ is that we know the distance between two triangles on
the surface.  This distance can either be measured in the intrinsic
metric (in which case it is trivial) or in the induced metric. For
comparison we tested both methods.

Let us first describe our algorithm for determining distances in the
induced metric at the discrete level.  Given two points on the
surface, we must find the shortest path, in the induced metric, along
the surface.  The problem here is that there are many definitions of
``geodesic'' which are equivalent in the continuum limit but differ at
the discrete level.
The algorithm we use is the following: given a triangle $t_0$ we find
the distance from its centre to the centre of all its neighbours. Then
we find the distance from those triangles to their neighbours
(excluding triangles already visited).  Iterating this procedure we
find the minimum distance to each triangle from $t_0$, subject to
the constraint that we have to pass through the centre of each 
triangle
traversed.  This is a piecewise linear approximation to the shortest
path, which should be good for sufficiently large paths.

Similarly, for the intrinsic metric, we can either define the distance
in units of jumps from triangle to triangle or, given that the
vertices have explicit $(\sigma_1,\sigma_2)$ coordinates in the
intrinsic formulation, we can define the shortest distance between
them as a straight line.

We have verified that these different definitions of distances give
identical results for the normal-normal correlation function measured
in the flat phase (modulo a trivial rescaling of the $r$-axis).  This
is illustrated in Fig.\ \ref{fig:geodes}.  The relevance of this
result goes beyond a simple consistency check: the fact that the
geodesics defined intrinsically and extrinsically coincide means 
also that intrinsic and the extrinsic metric overlap in the flat 
phase.

All of the normal-normal correlation functions presented in the paper
are obtained using method {\it A}.

\section*{Appendix C: Parallel Monte Carlo Algorithm}
\label{sec:para}
\begin{figure}
\centerline{ \epsfxsize=\floatwidth \epsfbox{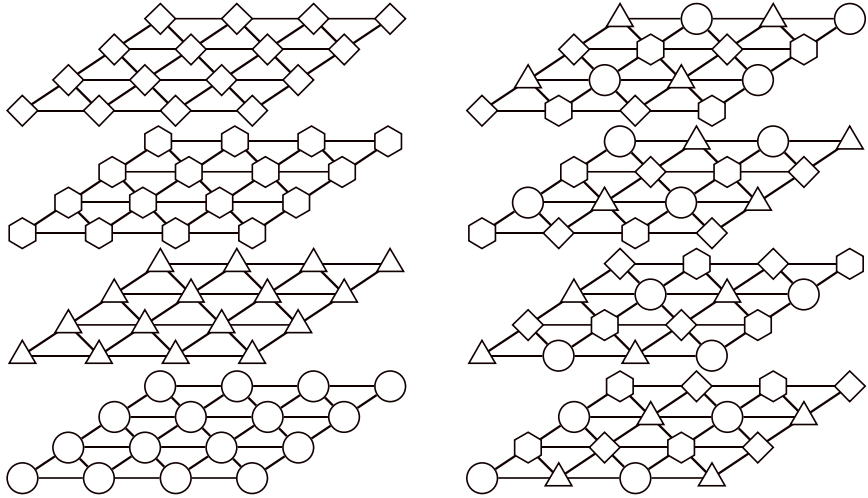}}
\caption{The improved parallelisation scheme.  On the left we show the
connectivity of the 4 different surfaces in the simulation, while on
the right we show the actual internal representation.  It is clear 
that,
on the right, no node ever has a neighbour of the same type.}
\label{fig:leaves}
\end{figure}
In this appendix we describe the parallel algorithm used on the MASPAR
MP1.  This machine is an old-style massively parallel computer with
16384 CPUs arranged in a $2-D$ ($128 \times 128$) mesh with nearest
neighbour connections.  Each individual CPU is a relatively small
processor (8 bit) with no floating-point unit.

A standard problem in using a parallel machine is the fact that the
amount of parallelisation, and consequently the performance increase,
is limited by the inter-dependence of the data.  In order to ensure
detailed balance in a Monte Carlo simulation, only a fraction of the
lattice can be updated concurrently.  In our case only 25\% of the
nodes can be updated. This translates into a huge performance loss as
75\% of the nodes remain idle.

In order to overcome this limitation we implemented an improved
parallelisation scheme.  Instead of simulating one surface we consider
4 independent Monte Carlo simulations.  The 4 corresponding meshes are
``interleaved'' in 4 arrays which store the node positions.  Each
surface is distributed onto the 4 arrays as shown in 
Fig.\ \ref{fig:leaves}.  The parallel machine updates one array at a 
time,
but now each array holds independent data and therefore all of it can
be updated in parallel.

We have compared the performance of this algorithm to the traditional
parallelisation and to the sequential code.  On the MASPAR, the
traditional parallelisation, i.e.\ a Monte Carlo simulation of a
single surface, can achieve a maximum speed of 80 Mflops (millions of
floating-point operations per second).  Our improved code is capable
of 280 Mflops, almost a four-fold increase.  This number is to be
compared with the peak performance of this machine which, measured
with the Linpack method, is around 440 Mflops.  Our scalar code on an
IBM RS/6000 390 with a clock of 66.5 MHz has a speed of 17 Mflops,
compared to a Linpack peak performance of 54 Mflops.

The parallel algorithm also requires a careful implementation of the
pseudo-random number generator.  While the intrinsic MASPAR random
number generator is known not to be reliable, using a sequential
random number generator would be incredibly time consuming.  The
solution to this problem is to have an independent random number
generator on each CPU, e.g.\ to regard it as an array-valued function.
In order to avoid cross correlations between the random sequences
generated by the parallel routine, a second standard random number
routine is used to seed the array.
\bibliography{paper}
\end{document}